\documentclass[pra,10pt,twocolumn,superscriptaddress,showpacs]{revtex4-2}

\usepackage[pdftex, pdftitle={Article}, pdfauthor={Author}]{hyperref}
\usepackage{epsfig}
\usepackage{amsmath}
\usepackage{graphicx}
\usepackage{graphics}
\usepackage{float}
\usepackage{amssymb}
\usepackage{natbib}
\usepackage{epstopdf}
\usepackage{wrapfig}
\usepackage{mathtools}
\usepackage{xcolor}
\RequirePackage{tcolorbox}
\RequirePackage[outline]{contour}
\usepackage{amsfonts}
\usepackage{subcaption}
\usepackage{mathtools}
\usepackage{dsfont}
\usepackage{bm}
\usepackage{physics}
\usepackage{verbatim}
\renewcommand{\bra}[1]{\left\langle #1\right|}
\renewcommand{\ket}[1]{\left| #1\right\rangle}
\renewcommand{\ip}[2]{\left\langle #1 | #2\right\rangle}

\renewcommand{\tr}{\text{Tr}}
\newcommand{\tot}[1]{#1_{\text{tot}}}

\usepackage[english]{babel}

\usepackage{hyperref}
 \hypersetup{
     colorlinks=true,
     linkcolor=blue,
     filecolor=magenta,      
     urlcolor=blue,
     citecolor=blue}
     
\usepackage{empheq}

\definecolor{fore}{RGB}{249,242,215}
\definecolor{myblue}{rgb}{.8, .8, 1}
\definecolor{forshading}{RGB}{185,217,255}
\newcommand*{\boxedcolor}{blue}
\renewcommand{\boxed}[1]{\textcolor{\boxedcolor}{\fbox{\normalcolor\m@th$\displaystyle#1$}}}
\makeatother

\graphicspath{{Figure/}}
\begin{document}

\title{The effect of non-selective measurement on the parameter estimation within spin-spin model}

\author{Ali Raza Mirza}
\affiliation{Department of Physics, University of Surrey, GU2 7XH, Guildford, United Kingdom}
\email{a.r.mirza@surrey.ac.uk}

\author{Jim Al-Khalili}
\affiliation{Department of Physics, University of Surrey, GU2 7XH, Guildford, United Kingdom}

\begin{abstract}
    \centering
\begin{tcolorbox}[colframe=white, colback=forshading, arc=10pt]
    We investigate the role of non-selective measurement on the estimation of system-environment parameters. Projective measurement is the popular method of initial state preparation which always prepares a pure state. However, in various physical situations of physical interest, this selective measurement becomes unrealistic. In this paper, we compare the estimation results obtained via projective measurement with the results obtained via unitary operation. We argue that in typical situations, parameters can be estimated with higher accuracy if the initial state is prepared with the unitary operator (a pulse). We consider the spin-spin model where a central two-level system (probe) interacts with the collections of two-level systems (bath). A probe interacts with a bath and attains a thermal equilibrium state, then via unitary operation, the initial state is prepared which evolves unitarily. The properties of the bath are imprinted on the reduced dynamics. Due to the initial probe-bath correlations present in the thermal equilibrium state, an additional factor arises in the dynamics which has a phenomenal role in the parameter estimation. In this paper, we study the estimation of bath temperature and probe-bath coupling strength which is quantified by the quantum Fisher information. Our results are promising as one can improve the precision of the estimates by orders of magnitude via non-selective measurement and by incorporating the effect of initial correlations.
\end{tcolorbox}
\end{abstract}
\maketitle
\section{Introduction}

Open quantum systems have attracted enormous attention because of their basic role in modern quantum technologies \cite{haroche2014exploring}. Since every quantum system interacts with its environment leading to decoherence \cite{schlosshauer2007decoherence, breuer2002theory}. The study of decoherence enables us to understand how we can harness quantum properties in the development and advancement of modern technologies \cite{schleich2016quantum}. One of the important quantum features is to sense information which is not possible with classical physics, known as quantum sensing \cite{degen2017quantum}. The key idea behind this is to utilize a quantum probe (a small controllable quantum system) undergoing decoherence \cite{benedetti2014quantum}. The use of probes allows us to extract some sensitive information about the environment. There are various theoretical tools available, one of them is to analytically derive the expression of quantum Fisher information (QFI) \cite{petz2011introduction}. This approach accredits not only the measurement but also quantifies the precision associated with it \cite{ather2021improving}. By incorporating the effect of initial correlation (present in the thermal equilibrium state), this precision can be enhanced by the order of magnitude \cite{mirza2024improving}.  Since the method of initial state preparation also influences the reduced dynamics, hence it is interesting to explore the impact of state preparation on the precision of estimates. By using the spin-spin model, we aim to investigate, if the initial state is prepared via unitary operation rather than conventional project measurement, how it affects quantum Fisher information, hence the estimation. Additionally, we incorporate the effect of initial correlations to explore further insights.

To learn about the bath parameters such as probe-bath coupling and bath temperature. The quantum probe interacts with its bath until they both attain an equilibrium state \cite{mirza2021master}. In due course, a suitable measurement is performed to prepare the probe in the desired initial state. The total prob-bath state evolves under the action of the total unitary operator. Studying the global probe-bath dynamics is quite challenging due to the large degrees of freedom of the bath. One possible way out is to use pure dephasing models \cite{morozov2012decoherence}. However, the drawback is that these models do not tell us anything about the energy exchange between the probe and the bath. Beyond the pure-dephasing, another choice is to use exactly solvable models such as the spin-spin model that considers $z-z$ interaction only \cite{majeed2019effect}. Once the dynamics are known, a measurement performed on the probe allows us to infer the bath properties such as temperature and coupling strength. A convenient parameter estimation approach is to determine quantum Fisher information that gives ultimate precision in our measurements \cite{chaudhry2015detecting}. According to the quantum Cramér-Rao bound, the variance in any unbiased parameter $x$ is bounded by the reciprocal of the Fisher information \cite{rao2008cramer}. Therefore, to maximise the precision in any estimator $x$, one has to maximise Fisher information over the interaction time.

To date, many attempts have been made to estimate parameters through quantum estimation theory. It is usual practice to consider the system and environment in a product state at $t=0$. Recent work, such as in Refs \cite{wu2020quantum, tamascelli2020quantum}, shows that the environment remains in a thermal equilibrium state all the time, and information about the bath is inferred through the quantum correlations established after state preparation. Within the harmonic oscillator bath, the single-qubit quantum probe has been utilised to estimate the cutoff frequency of bath oscillators \cite{benedetti2018quantum, benedetti2014characterization}. Squeezed probes have been subjected to investigation to improve the joint estimation of the nonlinear coupling and of the order of nonlinearity \cite{candeloro2021quantum}. On the other hand, using quantum resources, the sensitivity of phase estimation has been enhanced \cite{ciampini2016quantum}. However, these approaches disregard the quantum correlations that existed before the state preparation. Therefore, these findings are questionable, particularly when probe-bath coupling strength is strong. The initial probe-bath correlations present in the thermal equilibrium state have been extensively looked over \cite{chaudhry2013role,chaudhry2014effect,zhang2015role,chen2016effects,majeed2019effect}. More recently, \cite{mirza2024role, zhang2024improving} looked into the impact of these correlations in the parameter estimation via the Fisher information approach. Taking the basic seed of this idea, we extend our study to explore the effect of initial correlations in a spin environment and the effect of state preparation. Having a probe-bath thermal equilibrium state at hand, we start our analysis by preparing the probe's initial state via a unitary operation (a pulse). Then we work out the reduced dynamics of our probe. This would be essentially a $2\times2$ matrix which encapsulates the effect of unitary operation made to prepare the initial state, decoherence, and the initial correlations. In order to derive the expression of quantum Fisher information, we diagonalize this matrix and obtain eigenvalues and eigenvectors. The obtained Fisher information will be a function of the probe-bath interaction time and the estimator (temperature and coupling strength here). Then our goal is to optimise it over the interaction time such that QFI is maximum. We quantitatively show that initial correlations and state preparation can be manipulated to improve the accuracy of our measurements.

This paper is organised as follows: In the section \ref{model}, we model our quantum probe and bath with a paradigmatic spin-spin model and workout eigenstates. Then, in sec. \ref{dynamics},  we present the scheme of state preparations and the ensuing dynamics for both the cases with and without initial correlations. Next, In sec. \ref{fisher} we analytically derive an expression of quantum Fisher information and use it to estimate temperature (in \ref{tempo}) and probe-bath coupling strength (in \ref{couple}). Finally, we summarize our results in the section \ref{conclusion}.

\section{Spin-spin Model}\label{model}
We consider a single spin-half quantum system (probe) interacting with a bunch of spin-half quantum systems (bath). The total Hamiltonian can be written as
\begin{align}\label{eq1}
    H_{\text{tot}}
    =
\begin{cases}     
    H_{S0} + H_B + H_{SB} & \quad t\le 0,
\\ 
    H_{S} + H_B + H_{SB} & \quad t\ge 0,
\end{cases}   
\end{align}
where $H_{S0}$ is the system Hamiltonian before the system state preparation, with the parameters in $H_{S0}$ chosen to aid the state preparation process. $H_B$ is the bath Hamiltonian alone, and $H_{SB}$ is the system-bath interaction Hamiltonian. At $t = 0$, we prepare the initial state of our probe, and the system Hamiltonian becomes $H_S$ corresponding to its coherent evolution. Note that $H_{S0}$ is similar to $H_S$ in the sense that both operators live in the same Hilbert space, but they may have different parameters. Within spin-spin model, for $N$ spin-half systems in the bath,  we have (with $\hbar = 1$) 
\begin{subequations}\label{eq2}
\begin{eqnarray}
    H_{S0}
    &&= \frac{\varepsilon_0}{2} \sigma_z +\frac{\Delta}{2} \sigma_x;
\quad
    H_{S}
    = \frac{\varepsilon}{2} \sigma_z +\frac{\Delta}{2} \sigma_x,
\\
    H_{B}
    &&=\sum_{i=1}^{N} \left( \frac{\omega_{i}}{2} \sigma^{(i)}_z 
    + \chi_i \sigma^{(i)}_z \sigma^{(i+1)}_z \right),
\\
    H_{SB}
    &&= \frac{1}{2} \sigma_z \otimes g \sum_{i=1}^{N} \sigma^{(i)}_z. 
\end{eqnarray}
\end{subequations}
Here $\sigma_{x,y,z}$ are the Pauli spin operators, $\varepsilon_0$ and $\varepsilon$ denote the energy-level spacing of the central spin system before and after the state preparation respectively, $\Delta$ is the tunnelling amplitude, and $\omega_{i}$ denotes the energy level spacing for the $i^{\text{th}}$ spin in the bath. Bath spins interacts with each other via $\sum_{i=1}^{N} \chi_i \sigma^{(i)}_z \sigma^{(i+1)}_z$, where $\chi_i$ denotes the inter-spins interaction strength. Our probe interacts with the bath through interaction Hamiltonian $H_{SB}$, with $g$ is the probe-bath coupling strength. Note that our system Hamiltonian $H_S$ commutes with the total Hamiltonian meaning that the system energy is conserved. 

Now, our primary goal is to determine the dynamics of the probe. We express the interaction Hamiltonian into the system and bath operators as $H_{SB}=S\otimes B$, where $S$ is a system operator and $B$ is a bath operator. Now, the states $\ket{n}=\ket{n_1}\ket{n_2}\ket{n_3}...\ket{n_N} $ are the eigenstates of $B$, with $n_i=0, 1$ denoting the spin-up and spin-down along the $z$ axis respectively. We then have a set of eigenvalue equations
\begin{subequations}
\begin{eqnarray}
    g\sum_{i=1}^{N} \sigma^{(i)}_z\ket{n}
    &= {G}_{n} \ket{n},\label{eq3a} 
\\
    \sum_{i=1}^{N} \omega_{i} \sigma^{(i)}_z\ket{n}
    &= \Omega_{n} \ket{n}, \label{eq3b}
\\
    \sum_{i=1}^{N} \chi_i \sigma^{(i)}_z\sigma^{(i+1)}_z  \ket{n}
    &= \alpha_n \ket{n},\label{eq3c}
\end{eqnarray}
\end{subequations}
where $G_n= \sum^N_{i=1} (-1)^{n_i} g$, $\Omega_n= \sum^N_{i=1} (-1)^{n_i} \omega_i$, and $\alpha_n= \sum^N_{i=1}\chi_i (-1)^{n_i}(-1)^{n_{i+1}}$ are the eigenvalues of their respective operators. We also assume all environmental spins are coupled to the central spin with equal strength $g$.

\section{Initial state preparation and Dynamics}\label{dynamics}
Here we show analytical details of the initial state preparation process for both the cases with and without initial correlations. Then we show the calculations of the unitary operator and the evolution of both kind of initial states described below one by one.

\subsection{Without initial correlations}
We first discuss the preparation of the probe's initial state while correlations are ignored. In such a case, the probe and bath are initially in product state $\rho = \rho_{S0} \otimes \rho_{B}$, with $\rho_{S0} =e^{-\beta H_{S0}}/Z_{S0}$ and $ \rho_{B} = e^{-\beta H_B}/Z_B$ with the partition functions $Z_{S0} = \tr_S \left\{e^{-\beta H_{S0}}\right\}$ and $Z_B = \tr_B \left\{e^{-\beta H_B}\right\}$ where $\beta = 1/k_B T$. Note that this probe-bath state is only justified if the probe-bath interaction is weak enough. Under the sophisticated condition when $\varepsilon_0 \gg \Delta$, the probe state can be proven to be approximately `down' along the $z\text{-}$axis. Then, we make a suitable unitary operation to prepare the initial state.  For instance, if the desired probe's state is `up' along the $x\text{-}$axis, then an operator $R= e^{i \frac{\pi}{4} \sigma_y}$, realised by the application of a suitable control pulse, is implemented to the probe. The pulse duration is assumed to be sufficiently smaller than the effective Rabi frequency  $\sqrt{\varepsilon_0^2 + \Delta^2}$. After the pulse operation, we have
\begin{equation}
    \widetilde{\rho}_{\text{tot}}
     = \widetilde{\rho}_{S0} \otimes \rho_{B}\label{eq4}
\end{equation}
with $ \widetilde{\rho}_{S0} =  e^{-\beta \widetilde{H}_{S0}}/Z_{S0}$ and $\widetilde{H}_{S0} = R H_{S0} R^{\dagger}$. The action of pulse is represented by the `tilde' overhead the operators. Note that we can change the probe’s parameters as needed after the state preparation, that is, $\varepsilon_0 \rightarrow \varepsilon$. Doing so, the tunnelling term ($\frac{\Delta}{2} \sigma_x$) contributes significantly. Here, we assume shifting of parameters occurs within a very short time. Now the probe’s initial state can obtained by performing a trace over the bath. Thus we have (the superscript `u' stands for `uncorrelated initial state' since we are ignoring the probe-bath interaction) 
\begin{align*}
    \rho_{S0}^{u}
    =\frac{1}{Z_{S0}}\left\{\mathds{1}\cosh\left(\beta{\eta}_0\right) - \frac{\sinh\left(\beta{\eta}_0\right)}{{\eta}_0}\widetilde{H}_{S0}\right\},
\end{align*}
with ${\eta}_0 = (1/2) \sqrt{\varepsilon_0^2 + \Delta^2} $. It is useful to write this state in terms of components of the Bloch vector corresponding to this state
\begin{equation}\label{bloch1}
    \left( \begin{array}{c}
   n^{u}_x(0)\\
   n^{u}_y(0)\\
   n^{u}_z(0)
  \end{array} \right)
  	= \frac{\sinh\left(\beta{\eta}_0\right)}{Z_{S0}{\eta}_0}
	\left( {\begin{array}{c}
  	\varepsilon_0\\
   	0\\
   -\Delta\\
  \end{array} } \right).
\end{equation}
In order to make further progress, we need to determine reduced dynamics which necessitate the calculation of the total time evolution unitary operator first. This operator can be written as
\begin{align*}
    {U(t)
    =\sum_{n} e^{-i\frac{\Omega_{n}}2t}e^{-i\alpha_{n} t}e^{-i{H}^{n}_{S}t}\ket{n}\bra{n}
    =\sum_{n} U_{n}(t)\ket{n}\bra{n}},
\end{align*}
where $U_{n}(t) = e^{-i\frac{\Omega_{n}}2t}e^{-i\alpha_{n} t}e^{-i{H}^{n}_{S}t}$ which is only acting on the system's Hilbert space. Here ${H}^{n}_{S}
    \equiv \frac{{\xi}_{n}}{2} \sigma_z +\frac{\Delta}{2} \sigma_x$ is regarded as shifted Hamiltonian with energy parameter ${\xi}_{n} =  G_n + \varepsilon$. Now we can determine reduced density matrix as $\rho_u(t) = \text{Tr}_{B} \left[ U(t) \tot{\rho^u}(0) U^{\dagger}(t) \right]$  After some algebraic manipulations, we get
\begin{align} \label{dens2}
    \rho_{u}(t)
    &= \frac{1}{Z_B} \sum_{n}c_{n} U_{n}(t)\ket{\psi}\bra{\psi}  U^\dagger_{n}(t), \nonumber
\\
    &=\frac{1}{2}  
    \left( {\begin{array}{cc}
    1 +  {n^{u}_z(t)} & e^{-{\Gamma}_{u} (t)} e^{-i{\Omega_u}(t)} 
\\
    e^{-{\Gamma}_{u}(t)} e^{i{\Omega_u}(t)} & 1 -  {n^{u}_z(t)}
    \end{array} } \right),
\end{align}
where $Z_{B}=\sum_{n}{c_{n}}$, ${\Omega_u}(t) = \arctan\left[\frac{ {n^{u}_y(t)} }{ {n^{u}_x (t)}} \right]$, and the decoherence rate ${\Gamma}_{u}\left(t\right)
    = -\frac{1}{2}\ln \abs{ {\left\{n^{u}_{x}\left(t\right)\right\}^2} +  {\left\{n^{u}_{y}\left(t\right)\right\}^2}}$. Now the evolution of the Bloch vector components can be written in general form as 
\begin{align}
    {{n}^{u}_{i} (t)}
    = {\Theta^{u}_{ix}(t) {n}^{u}_{i} (0)}
\end{align}
with $i = x, y, z$, and the propagators are
\begin{subequations}\label{prop1}
\begin{eqnarray}
    {\Theta^u_{xx}(t)} 
    &=&\sum_{n}\frac{ c_{n}}{4 Z_B {\eta}^2_{n}}\Big\{\Delta^2 + {\eta}_{n}^{2}\cos(2{\eta}_{n}t)\Big\},
\\
    {\Theta^u_{yx}(t)}  
    &=&\sum_{n} \frac{c_{n} \varepsilon_n}{2 Z_B {\eta}_{n}}  \sin(2{\eta}_{n} t),
\\
    {\Theta^u_{zx}(t)} 
    &=&\sum_{n} \frac{c_{n} {\xi}_{n}\Delta}{2 Z_B {\eta}^2_{n}}\sin^2\left({\eta}_{n}t\right).
\end{eqnarray}
\end{subequations}
For convenience, we work in the dimensionless units, where every energy parameter is expressed in terms of $\varepsilon$. Thus, we have set $\hbar=k_{B} = 1$, during calculations and throughout the paper.

\subsection{With initial correlations}
To perceive the idea of an initial state containing initial correlations, we imagine that our probe has interacted with the bath to achieve a thermal equilibrium state; the Gibbs state $\rho_{\text{th}}= e^{-\beta H}/Z_{\text{tot}}$. Since the $\left[H_{S}, H_{SE}\right] \ne 0$, thus, in general, such a state can not be written as a product state. Our probe-bath state is, in general, a correlated state as the probe has interacted with the bath before. Now, we apply the same pulse which was used in the previous case to prepare the probe state. As a result, we have the correlated probe-bath state (the superscript `c' stands for `correlated state')
\begin{align}
    \rho^{c}_{\text{tot}}(0)
    &=\frac{1}{Z_{\text{tot}}} e^{-\beta \left(\widetilde{H}_{S0} + H_B + \widetilde{ H}_{SB}\right)},\label{eq5}
\end{align}
with $Z_{\text{tot}} = \tr_{SB} \left\{ e^{-\beta\left(\widetilde{H}_{S0} + H_B + \widetilde {H}_{SB}\right)} \right\}$ is the total partition function for the probe and the bath as a whole, and $\widetilde{H}_{SB} =  R H_{SB} R^{\dagger}$. Note that if the probe-bath interaction is sufficiently weak, this state would approximate the same product state given in Eq.~\eqref{eq4}. Looking at equations \eqref{eq3b} and \eqref{eq3c}, we can write $e^{-\beta H_B} \ket{n}
    = c_n \ket{n}$ with $c_n = e^{-\beta (\frac{\Omega_n}{2} + \alpha_n)}$. Also, we have
\begin{align}
    \left(\widetilde{H}_{S0} + \widetilde{H}_{SB}\right)\ket{n}
    &= \left(\frac{\varepsilon^n_{0}}{2} \sigma_z  -\frac{\Delta}{2} \sigma_x\right)\ket{n} \equiv H^n_{S0}\ket{n},\nonumber
\end{align}
where $H^n_{S0}$ is a `shifted' system Hamiltonian with a new energy parameter $\varepsilon_{0}^{n} = G_n + \varepsilon_0$. In this case, the Bloch vector components look like
\begin{equation}\label{bloch2}
    \left(
    \begin{array}{c}
        {n}^{c}_x (0)\\
        {n}^{c}_y (0)\\
        {n}^{c}_z (0)
    \end{array} \right)
  	     = \sum_n \frac{c_n\sinh\left(\beta{\eta}^n_0\right)}{Z_{\text{tot}}{\eta}^n_0}
	\left( {\begin{array}{c}
  	\varepsilon^n_0\\
   	0\\
   -\Delta\\
  \end{array} } \right),    
\end{equation}
now we have ${\eta}^n_0 = (1/2)\sqrt{(\varepsilon_0^{n})^2 + \Delta^2} $. Under the action of unitary operator, our probe-bath correlated initial state evolves. The reduced density matrix worked out to be
\begin{align}
    \rho_{c} (t)
    &= \frac{1}{\tot{Z}} \sum_{n} {J^{\text{corr}}_{n}} c_{n} U_{n}(t) \ket{\psi}\bra{\psi}U^\dagger_{n}(t),\nonumber
\\ 
    &=\frac{1}{2}  \label{dens1}
    \left( {\begin{array}{cc}
    1 +  {n^{c}_z(t)} & e^{-{\Gamma}_{c} (t)} e^{-i{\Omega_c}(t)} 
\\ 
    e^{-{\Gamma}_{c}(t)} e^{i{\Omega_c}(t)} & 1 -  {n^{c}_z(t)}
    \end{array} } \right),
\end{align}
where $\tot{Z} = \sum_{n}{J^{\text{corr}}_{n} c_{n}}$, $J^{\text{corr}}_{n} = 2\cosh(\beta \eta_{n})$, ${\eta}_{n} = (1/2)\sqrt{{\xi}_{n}^2 + \Delta^2}$, ${\Omega_c}(t) = \arctan\left[\frac{ {n^{c}_y(t)} }{ {n^{c}_x (t)}} \right]$, and the decoherence rate ${\Gamma}_{c}\left(t\right)
    = -\frac{1}{2}\ln \abs{ {\left\{n^{c}_{x}\left(t\right)\right\}^2} +  {\left\{n^{c}_{y}\left(t\right)\right\}^2}}$. The evolution of the corresponding Bloch vector components can be expressed in general form as
\begin{align}
    {n}^{c}_{i} (t)
    =\Theta^{c}_{ix}(t) {n}^{c}_{i} (0),
\end{align}
    with $i = x, y, z$. The propagators $\Theta^c_{ix} (t)$ are given as
\begin{subequations}\label{prop2}
\begin{eqnarray}
    \Theta^{c}_{xx}(t) 
    &=& \sum_{n}\frac{J^{\text{corr}}_{n} c_{n}}{4 Z_{\text{tot}} {\eta}^2_{n}}\Big\{\Delta^2 + {\eta}_{n}^{2}\cos(2{\eta}_{n}t)\Big\},
\\
    \Theta^{c}_{yx}(t)  
    &=& \sum_{n} \frac{J^{\text{corr}}_{n} c_{n} \varepsilon_n}{2 Z_{\text{tot}} {\eta}_{n}}  \sin(2{\eta}_{n} t),
\\
    \Theta^{c}_{zx}(t) 
    &=& \sum_{n} \frac{ J^{\text{corr}}_{n} c_{n} {\xi}_{n}\Delta}{2 Z_{\text{tot}} {\eta}^2_{n}}\sin^2\left({\eta}_{n}t\right).
\end{eqnarray}
\end{subequations}
If we compare these propagators with those given in \eqref{prop1}, we realise the replacement ${1}/{Z_B} \rightarrow {J^{\text{corr}}_{n}}/{\tot{Z}}$, which essentially captures the effect of initial correlations.

\section{Parameter estimation}\label{fisher}
The quantum Fisher information is related to the Cramer–Rao bond; the greater the QFI, maximum is the precision in our estimate. In this section, we first derive the formula of quantum Fisher information for our probe. Then we present estimation results in the subsequent sections.

\subsection{Quantum Fisher Information}
To quantify the precision with which a general environment parameter $\textit{x}$ (in our case this is temperature, $T$, or the coupling strength, $g$) can be estimated, we use quantum Fisher information which is defined by \cite{benedetti2018quantum}
\begin{align}
    {F}\left(x\right)
    &=\sum_{n=1}^2 \frac{(\rho'_n)^2}{\rho_n}+2\sum_{n\neq m }\frac{(\rho_n-\rho_m)^2}{\rho_n + \rho_m}\abs{\ip{v_m}{v'_n}}^2,\label{genfish1}
\end{align}
where $\rho_{m,n}$ and $v_{m,n}$ being eigenvalues and eigenvectors of any density matrix respectively. The superscript prime ( $'$ ) denotes the derivative with-respect-to the estimator \emph{x}. Thus, the first and foremost task is to diagonalize Eq. \eqref{dens2} and Eq. \eqref{dens1}. The eigenvalues of Eq. \eqref{dens1} are
    $\rho^c_{1}(t)
    = \frac{1}{2} \left[1 +  {\mathcal{N}_c(t)} \right]$, 
    $ \rho^c_{2}(t)
    = \frac{1}{2} \left[1 -  {\mathcal{N}_c(t)} \right]$
with
    $ {\mathcal{N}_c(t)}
    =\sqrt{ {\left\{n^{c}_{x}\left(t\right)\right\}^2} 
    +  {\left\{n^{c}_{y}\left(t\right)\right\}^2}
    +  {\left\{n^{c}_{z}\left(t\right)\right\}^2}}$. 
And corresponding eigenvectors are
\begin{subequations}
\begin{eqnarray}
    \ket{ {v^c_1}}
    &=\sqrt{\frac{ {\mathcal{N}_c} +  {n^{c}_z}}{2 {\mathcal{N}_c}}} \ket{\downarrow}_z
    - e^{-i{\Omega_c} } \sqrt{\frac{ {\mathcal{N}_c} -  {n^{c}_z}}{2 {\mathcal{N}_c}}} \ket{\uparrow}_z,
\\
    \ket{ {v^c_2}}
    &=\sqrt{\frac{ {\mathcal{N}_c} -  {n^{c}_z}}{2 {\mathcal{N}_c}}} \ket{\downarrow}_z
    + e^{-i{\Omega_c}} \sqrt{\frac{ {\mathcal{N}_c} +  {n^{c}_z}}{2 {\mathcal{N}_c}}} \ket{\uparrow}_z,
\end{eqnarray}\label{eigenvec}
\end{subequations}
where $\ket{\uparrow}_z$ and $\ket{\downarrow}_z$ are eigenstates of $\sigma_z$ with eigenvalues $+1$ and $-1$ respectively. If we disregard initial correlations, we obtain a similar set of eigenvalues and eigenvectors but having superscript `u' with $J^{\text{corr}}_n = 1$. Now we are equipped to write the final expression of quantum Fisher information, taking initial correlations into account. We have
\begin{align}\label{fish1}
    {F}_{c}
    &=\frac{\left(\Gamma'_{c} - {n}^{c}_z \left({n}^{c}_z\right)' e^{2{\Gamma_c}}\right)^2}{{f}_c \left(e^{2 {\Gamma}_c} - {f}_c\right)}
    + \frac{\left(\left({n}^{c}_z\right)' + {n}^{c}_z {\Gamma'_c} \right)^2}{{f}_c}
    + \frac{\left(\chi'_{c} \right)^2}{e^{2  {\Gamma_c}}},
\end{align}
with ${f}_c
    =1 + \left({n}^{c}_z\right)^2 e^{2 {\Gamma}_c}$. In the chosen model, both diagonal and off-diagonal entries evolve. Therefore, we can see the Fisher information also depends on the time-dependent factor ${n}^{c}_z$ unlike the pure-dephasing case where only off-diagonal entries evolve. If we set ${n}^{c}_z = 0$, implying ${f}_c = 1$, hence we recover the Fisher information given in Ref. \cite{ather2021improving}, benchmarks our calculations. If  initial correlations are discarded, QFI is then 
\begin{align}
    {F}_{u}\label{fish2}
    &=\frac{\left(\Gamma'_{u} - {n}^{u}_z \left({n}^{u}_z\right)' e^{2{\Gamma_u}}\right)^2}{{f}_u \left(e^{2 {\Gamma}_u} - {f}_u\right)}
    + \frac{\left(\left({n}^{u}_z\right)' + {n}^{u}_z {\Gamma'_u} \right)^2}{{f}_u}
    + \frac{\left(\chi'_{u} \right)^2}{e^{2  {\Gamma_u}}},
\end{align}
with ${f}_u
    =1 + \left({n}^{u}_z\right)^2 e^{2 {\Gamma}_u}$.
\subsection{Estimating environment temperature}\label{tempo}
\begin{figure}[t]
    \includegraphics[scale=0.93,trim={0cm 0cm 0cm 0cm}]{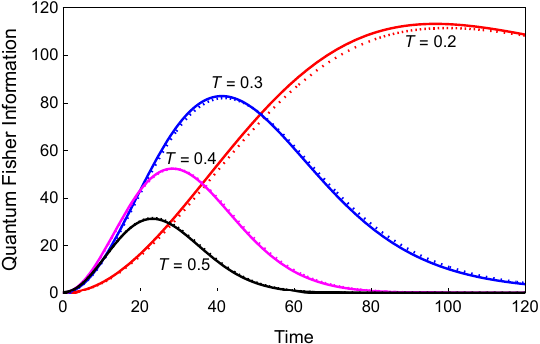}
    \caption{(Color online) The behaviour of QFI as a function of time while estimating temperature. Solid curves include the correlations' effect whereas dotted ones ignore this effect. The number of environmental spins is $N=50$, coupling strength $g = 0.01$, and inter-spin interaction $\chi=0$. The rest of probe-bath parameters are $\omega_i = 1, \varepsilon_0= 4, \varepsilon = 2$ and $\Delta = 1$.}
	\label{f1}
\end{figure}
\begin{figure}[t]
    \includegraphics[scale=0.93,trim={0cm 0cm 0cm 0cm}]{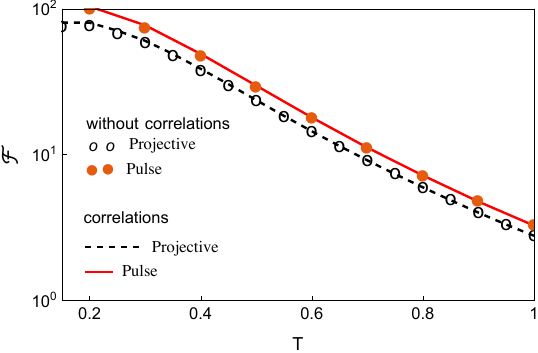}
    \caption{(Color online) The behaviour of optimised QFI, that is, $\mathcal{F}$ estimating temperature $T$. The rest of the probe-bath parameters are the same as Fig. \ref{f1}}
    \label{f2}
\end{figure}
\begin{figure}[t]
    \includegraphics[scale=0.93,trim={0cm 0cm 0cm 0cm}]{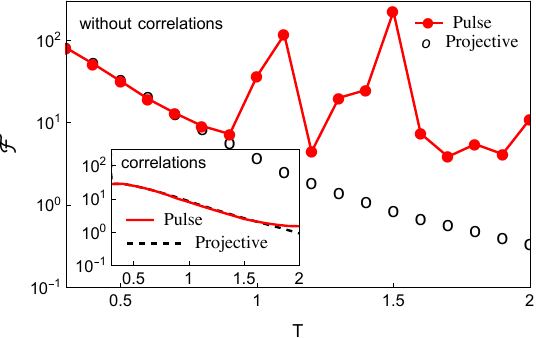}
    \caption{(Color online) Same as Fig. \ref{f1} except that now we have stronger coupling strength $g=1$.}
    \label{f3}
\end{figure}
Having all these analytics at hand, we can now move to the main part of this paper which relies on the results of estimation. Recall, that the primary goal here is to investigate the role of initial correlations and state preparation to look for maximum Fisher information. Our QFI is a function of time, temperature and probe-bath coupling strength. To estimate bath temperature with ultimate precision, we need to find the interaction time such that QFI is maximum. To proceed, we first need to calculate partial derivatives with-respect-to temperature $T$ and use them in Eqs. \eqref{fish1} and \eqref{fish2}. The effect of correlations is encapsulated by the factor $J^{\text{corr}}_n$ appearing in the propagators $\Theta^c_{ix} (t)$. First, we consider probe-bath coupling to be weak where the effect of correlations is expected to be less \cite{mirza2024role, chaudhry2013role, mirza2023improving}, which in turn negligible impacts the accuracy. Fig. \ref{f1} shows the behaviour of quantum Fisher information as a function of time at various temperatures. The solid curves signify QFI taking initial correlations into account whereas dotted curves discard the effect of correlations. Peak values represent the optimised QFI which is in turn the ultimate precision in the temperature estimation. Here we consider the non-interacting ($\chi=0$) spins in the bath $N=50$. We see the effect of the initial correlation is almost negligible, as the coupling strength is very small $g=0.01$. Next, we notice that the peak is maximum at lower temperatures which means low temperature is favourable for better estimation. Since a quantum state is very sensitive to the temperature; as the temperature is raised, the decoherence process speeds up (the loss of quantum properties), hence the precision decreases with temperature.
\begin{figure}[t]
    \centering
    \includegraphics[scale=0.93,trim={0cm 0cm 0cm 0cm}]{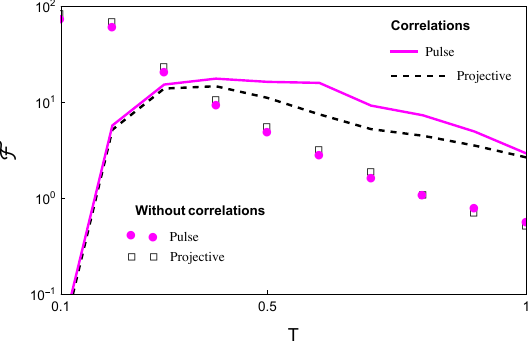}
    \caption{(Color online) Same as Fig. \ref{f1} except that now we have incorporating the inter-spin interaction $\chi = 0.1$ in smaller environment $N=10$.}
    \label{f4}
\end{figure}
In the rest of all figures from now, we compare our results with the results of Ref. \cite{mirza2024role} where the initial state was prepared via usual projective measurement. Under the same set of probe-bath parameters as in Fig. \ref{f1}, we show the behaviour of optimised quantum Fisher information as a function of temperature $T$, for the case if the initial state is prepared via projective measurement (call it $\mathcal{F}_{\text{proj}}$ in black-dashed), versus $\mathcal{F}_{\text{pulse}}$ (in red-solid if the initial state is prepared via unitary operator) in Fig. \ref{f2}. We notice that $\mathcal{F}_{\text{pulse}}$ is slightly higher than $\mathcal{F}_{\text{proj}}$, if initial correlations are incorporated. Here comes the unequivocal advantage of considering non-selective measurement rather than projective. We repeat this for both cases but without initial correlations using Eq. \eqref{fish2}, nevertheless same behaviour is seen as solid circles (pulse) and empty circles (projective) overlap with their respective curves of correlations. The reason is obvious that within a weak coupling regime, the correlation energy is dominated by as thermal energy $\beta$, as always \cite{mirza2024role}. 

However, if coupling strength increased to $g=1$, the difference between $\mathcal{F}_{\text{pulse}}$ and $\mathcal{F}_{\text{proj}}$ amplifies if we discard initial correlations. As shown in main Fig. \ref{f3}, $\mathcal{F}_{\text{pulse}}$ (solid-circled-red) is greater than $\mathcal{F}_{\text{proj}}$ (black-empty circles) for higher values of temperature, which is what we were expecting. If we notice, the Bloch vector components are given in Eqs. \eqref{bloch1} and \eqref{bloch2}, explicitly depend on temperature. As temperature increases, the orientation of the initial state changes, that is, \emph{x} and \emph{y} components of the Bloch vector decrease in magnitude, as a result, the degree of mixedness increases. This improves the precision of temperature estimation by the order of magnitude. However, if we take the initial correlations into account, $\mathcal{F}_{\text{pulse}}$ (solid-red) and $\mathcal{F}_{\text{proj}}$ (dashed-black) almost overlap as shown in the inset of figure. This is because thermal energy and interaction energy equally dominate, hence the effect of state preparation almost disappears. As a final comment; higher temperature and non-selective measurement (with a pulse) are favourable as they give ultimate accuracy.

Next, we investigate the impact of inter-spin interaction $\chi=0.1$. With a small number of bath spins, the decoherence process slows down, as a result, initial correlations and the role of state preparation can be better realised. Results are shown in Fig. \ref{f4} in a smaller bath with $N=10$. Figure compare the behaviour of $\mathcal{F}_{\text{pulse}}$ (solid-magenta) versus  $\mathcal{F}_{\text{proj}}$ (dashed-black), if correlations are considered. One can witness that non-selective measurements made at $t = 0$, produce larger QFI than QFI achievable with projective measurement. On the other hand, in uncorrelated cases, no appreciable role of state preparation has been seen. In the smaller spin bath, we expected more Fisher information than in the larger bath. However, we notice that $\mathcal{F}_{\text{plulse}}$ with $N=10$ in Fig. \ref{f4} is less than $\mathcal{F}_{\text{plulse}}$ with $N=50$ [Fig. \ref{f3}]. This means that inter-spin interactions have played a significantly negative role in precision improvement as $\mathcal{F}_{\text{plulse}}$ has been suppressed in Fig. \ref{f4}. Conclusively, inter-spin interaction has to be kept minimum to improve the accuracy which can be done by keeping bath spins at a distance from each other.

\subsection{Estimating probe-bath coupling strength}\label{couple}
Next, we consider the impact of state preparation on the estimation of coupling strength. Again, we consult with Eq. \eqref{fish1} and Eq. \eqref{fish2} but this time we need derivatives with-respect-to coupling strength $g$. 
\begin{figure}[h]
    \centering
    \includegraphics[scale=0.93,trim={0cm 0cm 0cm 0cm}]{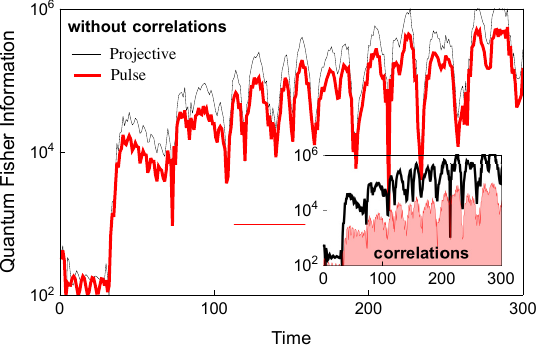}
    \caption{(Color online) The behaviour of quantum Fisher information while estimating coupling strength $g=0.1$ at temperature $T=1$. The rest of the probe-bath parameters are the same as Fig. \ref{f1}}
    \label{f5}
\end{figure}
\begin{figure}[h]
    \centering
    \includegraphics[scale=0.93,trim={0cm 0cm 0cm 0cm}]{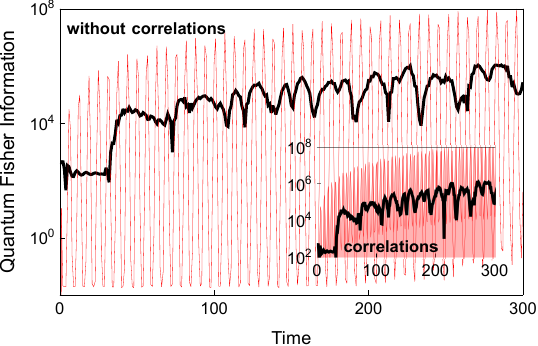}
    \caption{(Color online) Same as Fig. \ref{f5} except that now we have $g=0.5$ and $T=0.5$.}
    \label{f6}
\end{figure}
Results are illustrated in Fig. \ref{f5}, where we have shown the QFI as a function of interaction time, keeping temperature and coupling strength to be fixed at $T=1$ and  $g=0.1$ respectively. Red-solid curves denotes $\mathcal{F}_{\text{pulse}}$ whereas black curves signify $\mathcal{F}_{\text{proj}}$. At least two comments can be made regarding this result. First, the quantum Fisher information keeps on increasing as a function of time unlike in the case of coupling strength estimation where we get peaks, like in Fig. \ref{f1}. The source of this continuous increase is obviously the derivatives $\frac{\partial}{\partial_g} \Gamma$, $\frac{\partial}{\partial_g} \Omega$ and $\frac{\partial}{\partial_g} n_z$ which oscillate very fast in the long time limit. Therefore, our measurement result becomes extremely sensitive to the coupling strength $g$. Thus the interaction time becomes the subject of what the order of accuracy one require. The same behaviour has been seen in the Ref. \cite{mirza2024role}. Secondly, if we ignore correlations, $\mathcal{F}_{\text{proj}} >  \mathcal{F}_{\text{pulse}}$ all the times. A similar trend prevails if we incorporate the effect of correlations as shown in the inset. It means at higher temperatures, while estimating the coupling strength, the projective measurement method is favourable than the pulsed one which is under the consideration. However, the situation drastically changes if we jump into the low-temperature regime. Fig. \ref{f6} depicts the behaviour of quantum Fisher information as a function of time at a fixed value of temperature $T=0.5$ and coupling strength $g=0.5$. In either with or without correlation case, $\mathcal{F}_{\text{pulse}}>\mathcal{F}_{\text{proj}}$ all the times. The objective of our work is once again quite clear as one can clearly witness that higher accuracy can be availed if non-selective measurement is performed rather than projective.

\section*{Conclusion}\label{conclusion}
Initial correlations and non-selective measurement are the two basic elements in our analysis. Decoherence is the challenge for both of these. We considered a variety of physical situations and investigated how to get the best estimates using a quantum probe. Our study revealed how to choose an engineer bath or choose temperature such that error in our measurement is minimal. Results presented in this paper divulged that the role of initial state preparation and initial correlations can be very significant, especially in the strong coupling regime and at low temperatures. This entailed a remarkable impact on quantum sensing as we saw one can get ultimate precision in the estimates via non-selective measurement and by incorporating the effect of initial correlations. In conclusion, our results give an overview of how precision is linked with other probe-bath parameters.

\section*{acknowledgements}
A.~R.~Mirza and J. Al-Khalili are grateful for support under the grant RN0491A from the John Templeton Foundation Trust.

\bibliographystyle{unsrt}
\bibliography{Library}

\end{document}